\documentclass{PoS}

\title{Fisher's Zeros and Perturbative Series in Gluodynamics}
\ShortTitle{Fisher's Zeros and Perturbative Series}

\author{A. Denbleyker, D. Du, and  \speaker{Y. Meurice}\\
Department of Physics and Astronomy, The University of Iowa,
Iowa City, Iowa 52242, USA\\
E-mail: \email{alan-denbleyker@uiowa.edu}\\
E-mail: \email{daping-du@uiowa.edu}\\
E-mail:\email{yannick-meurice@uiowa.edu}}
\author{A. Velytsky\thanks{Current address: 
EFI, University of Chicago, 5640 S. Ellis Ave., Chicago, IL 60637
and Argonne National Laboratory, 9700 Cass Ave., Argonne, IL 60439} \\
Department of Physics and Astronomy, UCLA, Los Angeles, CA 90095-1547, USA\\
E-mail: \email{vel@physics.ucla.edu}}     

\abstract{We study the zeros of the partition function in the complex $\beta$ plane 
(Fisher's zeros) in SU(2) and SU(3) gluodynamics. We discuss their effects on the asymptotic
behavior of the perturbative series for the average plaquette. We present
new methods to infer the existence of these zeros in region of the complex $\beta$ plane where MC reweighting
is not reliable. These methods are based on the assumption that the plaquette distribution 
can be approximated by a $\phi^4$ type distribution. 
We give new estimates of the locations for a $4^4$ lattice.
For $SU(2)$, we found zeros at $\beta =2.18(1) \pm i0.18(2)$ (which differs from previous estimates), and at 
$\beta =2.18(1) \pm i0.22(2)$.  
For $SU(3)$, we confirm $\beta =5.54(2)\pm i0.10(2)$ and found additional 
zeros at 
$\beta =5.54(2)\pm i0.16(2)$. Some of the technical material can be found in recent preprints, in the 
following we emphasize the motivations (why it is important to know the locations of the zeros) and 
the challenges (why it is difficult to locate the zeros when the volume increases).  
}

\FullConference{The XXV International Symposium on Lattice Field Theory\\
		 July 30-4 August 2007\\
		 Regensburg, Germany}

\begin{document}

\section{Introduction}
More than 200 years ago, L. Euler \cite{euler} realized that the asymptotic behavior of the expansions coefficients of the Bessel functions about the origin could be obtained from the location of the complex zeros. 
For a single plaquette with $SU(2)$ gauge group (as defined, for instance, in Ref. \cite{plaquette}, as usual $\beta=4/g^2$), we obtain that in this simple model the average plaquette $P$ can be written as 
\begin{equation}
P=1-2\beta\sum_{n=1}^{\infty}\frac{1}{\beta^2+\alpha_n^2}\ ,
\end{equation}
where the $\alpha_n$ are the locations of the zeros of $J_1(z)$ (on the positive real axis).
Using the geometrical series one sees that the radius of convergence of the expansion in $\beta$ is $1/\alpha _1$. This result can be rephrased in the following way: the radius of convergence of the 
strong coupling expansion of $P$ is the distance from the origin to the closest zero of the partition function (which in this simple example is located, as well as all the other zeros, on the imaginary axis). 

The situation is more interesting at large $\beta$ and fundamentally different 
from the case of scalar models. As well known, the zeros of $J_1(z)$ become approximately equally spaced for large real argument. If we now consider $P$ in the  $g^2=4/\beta$ complex plane, the zeros accumulate at zero 
along the imaginary axis which plays the role of a Stokes line. Unlike the scalar case the partition function is well defined at real negative $g^2$, but as we cross the imaginary axis, the values of 
$P$ changes discontinuously. This can be seen from the sum rule $P(\beta)+P(-\beta)=2$, which remains  true on arbitrary lattices with even number of sites in every directions \cite{gluodyn04}. 

We expect that the features of the strong and weak coupling expansions observed for the one plaquette 
model to remain true on generic lattices. However, other zeros of the partition function related to phase transitions are expected. For the 2-dimensional Ising model, it has been observed by M. 
Fisher \cite{fisher} that the zeros of the partition function pinch the real axis of the 
complex inverse temperature at $\beta_c$.  For the 3-dimensional Ising model, numerical calculations  consistent with this scenario have 
been obtained \cite{pearson} on a $4^3$ lattice. These zeros in the complex $\beta$ plane are called 
Fisher's zeros and should be distinguished from the Lee-Yang zeros in the complex magnetic field plane. 

For a zero temperature lattice gauge theory with a Wilson action, we expect no phase transition 
on the real $\beta$ axis. However, such theories can be seen as ``close'' to other theories (e. g. 
at non-zero temperature or with a positive adjoint coupling) that have a phase transition.  
Consequently, it is plausible that zero temperature lattice gauge theory with a Wilson action have 
Fisher zeros close to the real axis but that these zeros do not pinch the real axis in the 
infinite volume limit. This is a plausible explanation \cite{third} for the unexpected behavior \cite{rakow2002} of the weak coupling 
expansion of $P$ for $SU(3)$ \cite{direnzo2000,rakow05}. 
Standard methods of series analysis suggest \cite{rakow2002,third}  
a singularity on the real axis, namely $P\propto (1/5.74-1/\beta)^{1.08}$. 
This would imply a peak in the second derivative of $P$ with a height increasing with the 
volume, which is not seen at zero temperature \cite{third}. 
The vicinity of the critical point in the fundamental-adjoint plane, suggests the 
approximate mean field behavior \cite{third}:
\begin{equation}
-\partial P/\partial \beta \propto {\rm ln}((1/\beta_m-1/\beta)^2+\Gamma^2)\ ,
\label{eq:mean}
\end{equation}
Fits of the series with such parametric form yield the approximate values $\beta_m\simeq 5.78$ and 
$\Gamma \simeq 0.006$ (i.e $Im \ \beta \simeq 0.2$). 
 
Values of $\Gamma$ which are too large (too small) would produce modulations of the coefficients 
(peaks in the derivatives of $P$) which are not observed. 
A detailed analysis \cite{third} yield the bounds (for $SU(3)$) $
0.001<\Gamma<0.01$. 
This suggests zeroes of the partition function in the complex $\beta$ plane with 
\begin{equation}
0.03\simeq0.001\beta_m^2<Im \beta <0.01\beta_m^2\simeq 0.33
\end{equation}

In these proceedings, we report recent efforts to locate such zeros in pure gauge theories with 
gauge group $SU(2)$ and $SU(3)$. Some of the technical details can be found in a recent preprint \cite{quasig}. Another preprint is in preparation \cite{fisherw} and should be available soon.  
In order to avoid repetitions, we will emphasize the motivations and aspects not covered in these 
preprints.

\section{The non-perturbative part of the plaquette and the gluon condensate}
It is quite common that the difference between a physical quantity and its perturbative 
expansion is of the form $\exp(-K/g^2)$. One of the best known example is the quantum 
mechanical double-well where the perturbative series is not able to take into account the 
tunneling effect and instantons are needed. For the average plaquette, the issue 
is obscured by the hypothetical zero close to the real axis and the factorial growth 
necessary to get an envelope \cite{npp} in the accuracy versus coupling at successive order is not 
reached at the order where the perturbative expansion is available. Larger order extrapolation are necessary. Two models have been considered \cite{npp}.

The first is based on Eq. (\ref{eq:mean}) which implies a  dilogarithmic series for $P$:
\begin{equation}
P\sim \sum _{k=0} a_k\beta^{-k}\simeq C({\rm Li}_2 (\beta^{-1}/(\beta_m^{-1}+i\Gamma))+{\rm h.c}\ .
\label{eq:dilog}
\end{equation}
The low order coefficients depend very little on $\Gamma$ as long as the upper bound given above is 
satisfied. 
It is remarkable that the method of stochastic perturbation theory allows us to calculate the series up to 
order 10 \cite{direnzo2000} and 16 \cite{rakow05}, however 
larger series are needed to resolve $\Gamma$! As we will explain below, the situation might be better for $SU(2)$. The two other parameters can be determined using for instance the values of $a_9$ and $a_{10}$.  This gives 
very good predictions \cite{npp} of the values of $a_8, a_7, \dots $. This nice 
regularity is still begging for a diagrammatic explanation.

Despite its predictive success, the dilogarithmic series has a finite radius of convergence and the 
coefficients do not have the expected factorial growth that is observed in the one plaquette model. 
The second extrapolation was based on an IR renormalon model \cite{mueller93,itep,burgio97}
\begin{equation}
\sum _{k=0} a_k\bar{\beta}^{-k} \simeq K \int_{0}^{\infty}dt {\rm e}^{-\bar{\beta}t}\ (1-t\ 33/16\pi^2)^{-1-204/121}
\end{equation}
\begin{equation}
\bar{\beta}=\beta(1+d_1/\beta+\dots)
\label{eq:shift}
\end{equation}

These two extrapolations seem consistent with the behavior
\begin{equation}
P(\beta)-P_{pert.}(\beta) \simeq C (a/r_0)^4
\label{eq:nppl}
\end{equation}
with $a(\beta)$ defined with the force scale \cite{force01,force98} with $r_0=0.5$ fm, and $P_{pert}$ appropriately truncated. For large $\beta$ this has the desired exponential form. 
Attempts have been made in the past \cite{dig81,rakow2002,rakow05} to relate $C$ to the so-called gluon condensate \cite{itep}. Remembering the $\alpha/\pi$ factor in the definition, the value that could in principle be compared with the commonly used value of 0.012 $GeV^4$ is $(36/\pi^2)Cr_0^{-4}$ for $N_c=3$. $C$ is sensitive to resummation. $C\simeq 0.6$ with the bare series \cite{npp} 
and 0.4 with the tadpole improved series \cite{rakow05}. This gives 
values 3-5 times larger than the value quoted above. Besides the question of scheme dependence, the gluon condensate is not an order parameter and it seems difficult to compare the lattice 
results with quantities defined in the context of sum rules. On the other hand, it is important 
to figure out how well the scaling with the lattice spacing given in Eq. (\ref{eq:nppl}) is obeyed 
and if it can be explained semi-classically as it can be done for the double-well. 
In this context, it is crucial to understand the complex singularities of $P$, which complicate 
the analysis of the scaling. 

\section{The zeros of the partition function}
\noindent
The zeros of the partition function can be located with the reweighting method \cite{falcioni81,alves}.
\begin{equation}
Z(\beta_0+\Delta \beta)=Z(\beta_0)<\exp (-\Delta \beta S)>_{\beta_0}\ .
	\label{eq:pf}
\end{equation}
It is convenient to subtract $<S>$ from $S$ in the exponential because it removes fast 
oscillations without changing the complex zeroes. 
$Z(\beta)$ is the Laplace transform of density of states $n(S)$:
\begin{equation}
	Z(\beta)=\int_0^{\infty}dS\ n(S)\exp(-\beta S) \ .
\end{equation}
If we could estimate $Z(\beta)$ along an axis in the imaginary direction, we could calculate the 
inverse Laplace transform and obtain $n(S)$. 
\begin{figure}
\centerline{\includegraphics[angle=270,width=3in]{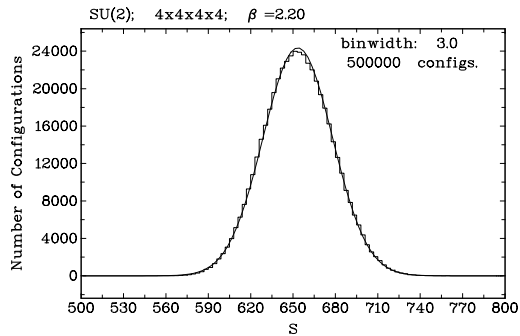}}
\caption{Distribution of the 500,000 values of $S$ in an histogram with 100 bins 
for a $SU(2)$ pure gauge theory on a $4^4$ lattice at $\beta =2.20$. The solid line is a Gaussian fit.}
\label{fig:su2hist}
\end{figure}
As shown in Fig. \ref{fig:su2hist}, the distribution of values of $S$ for $SU(2)$ can be fitted very well with a Gaussian with $\sigma_S^2=<S^2>-<S>^2$. 
This suggests \cite{alves} a criterion to determine a region of confidence for MC 
zeros. For a Gaussian distribution, the fluctuation in $\exp (-\Delta\beta (S-<S>))$ 
is smaller than the average for $|\Delta\beta|^2< \ln (N_{conf.})/\sigma_S^2$.
This defines a radius of confidence {\it $\sqrt{\ln (N_{conf.})}/\sigma_S$} 
in the complex $\beta $ plane. As $\sigma_S^2\propto V$, the radius of confidence shrinks like $V^{-1/2}$  which is hard to beat with $\sqrt{\ln N_{conf.}}$. 

When the zeros pinch the real axis 
as the volume increases, which is expected near a phase transition, there is some hope that it is possible to follow them with MC methods. This explains some positive results \cite{alves} on $2\times L^3$ and $4\times L^3$ lattices. On the other hand, on $L^4$ lattices, we expect the zeros to stabilize away from the real axis and MC methods can only give lower bounds that shrink when $L$ increases. 
For instance for $SU(3)$ on a $8^4$ lattice \cite{latt06}, we were only able to verify that 
$Im \beta >0.03$. In addition, for $SU(2)$, there seem to be no zeros in the Gaussian region of 
confidence even on a $4^4$ lattice. For these reasons we have developed new methods to find zeros of the partition that lay outside of the region of confidence of MC calculations. 
\section{New methods to locate the zeros \cite{quasig,fisherw}}

Gaussian distributions (of $S$) have no complex zeros.
The Gaussian circle of confidence in the complex $\beta$ plane defined by the condition 
$\sigma _f<|f|\sqrt{N_{conf.}}$. If this criterion is applied directly to a non-Gaussian 
distribution having complex zeros, it will automatically exclude the regions that contain the 
zeros. When looking for complex zeros, we look for the intersection of the zero level curves for 
the real and imaginary parts. We are interested in knowing how much these level curves can move due 
to statistical fluctuations. 
We proposed \cite{quasig} to consider the alternative region of confidence defined by a 
condition that controls the error on the level curves:
\begin{equation}
\sigma _f <d {\sqrt{N_{conf.}}\ |f'|}\ .
\label{eq:newconf}
\end{equation}
In order to be useful $d$ should be a fraction of the typical distance between zero level curves of the real and imaginary part. This methods has allowed \cite{quasig} to reject dubious zeros on the edge of the Gaussian 
circle of confidence for $SU(2)$ on a $4^4$ lattice. It remains applicable when the deviation from a Gaussian distribution is significant and true zeros appear. 

\begin{figure}
\includegraphics[width=2in,angle=270]{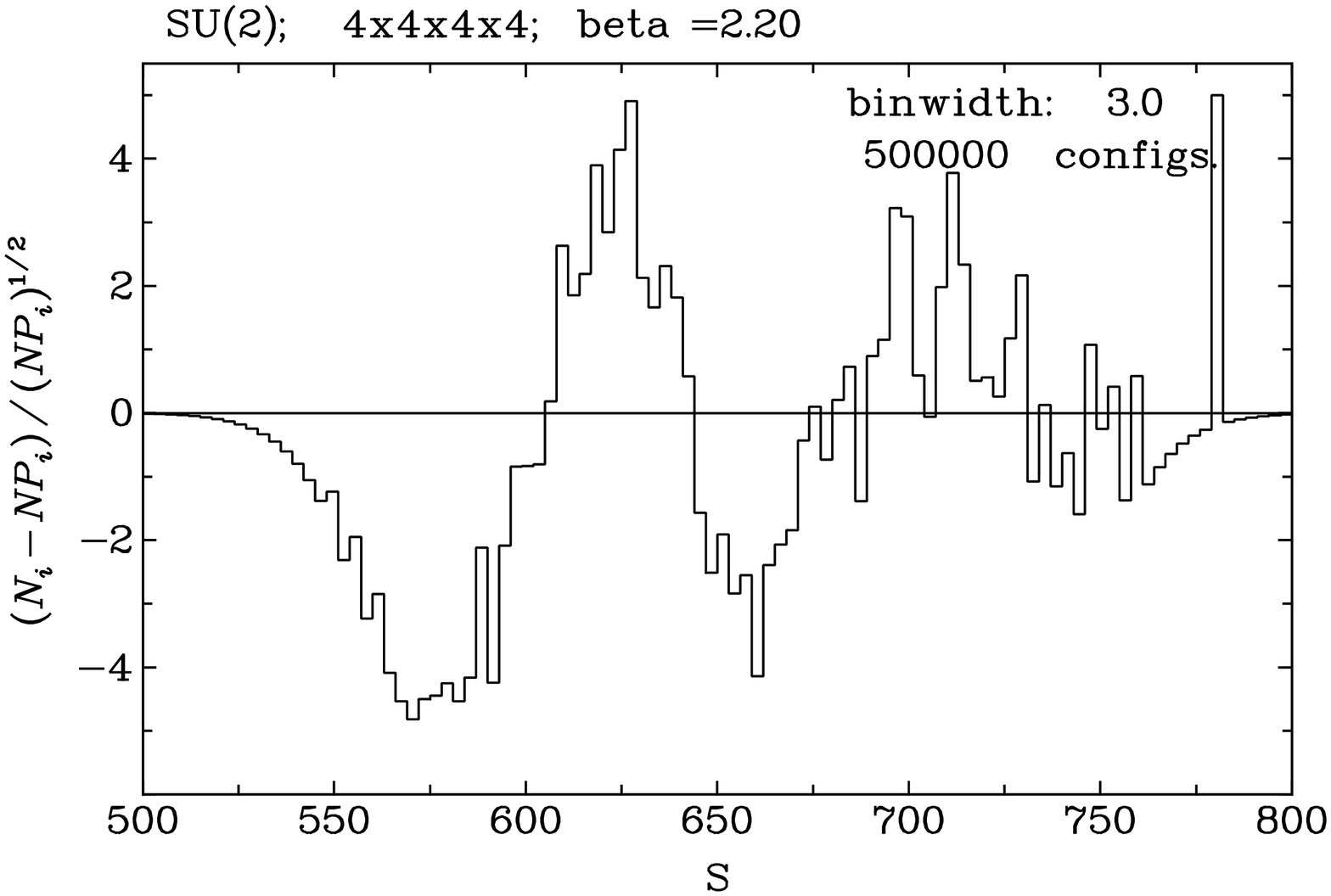}
\includegraphics[width=2.in,angle=270]{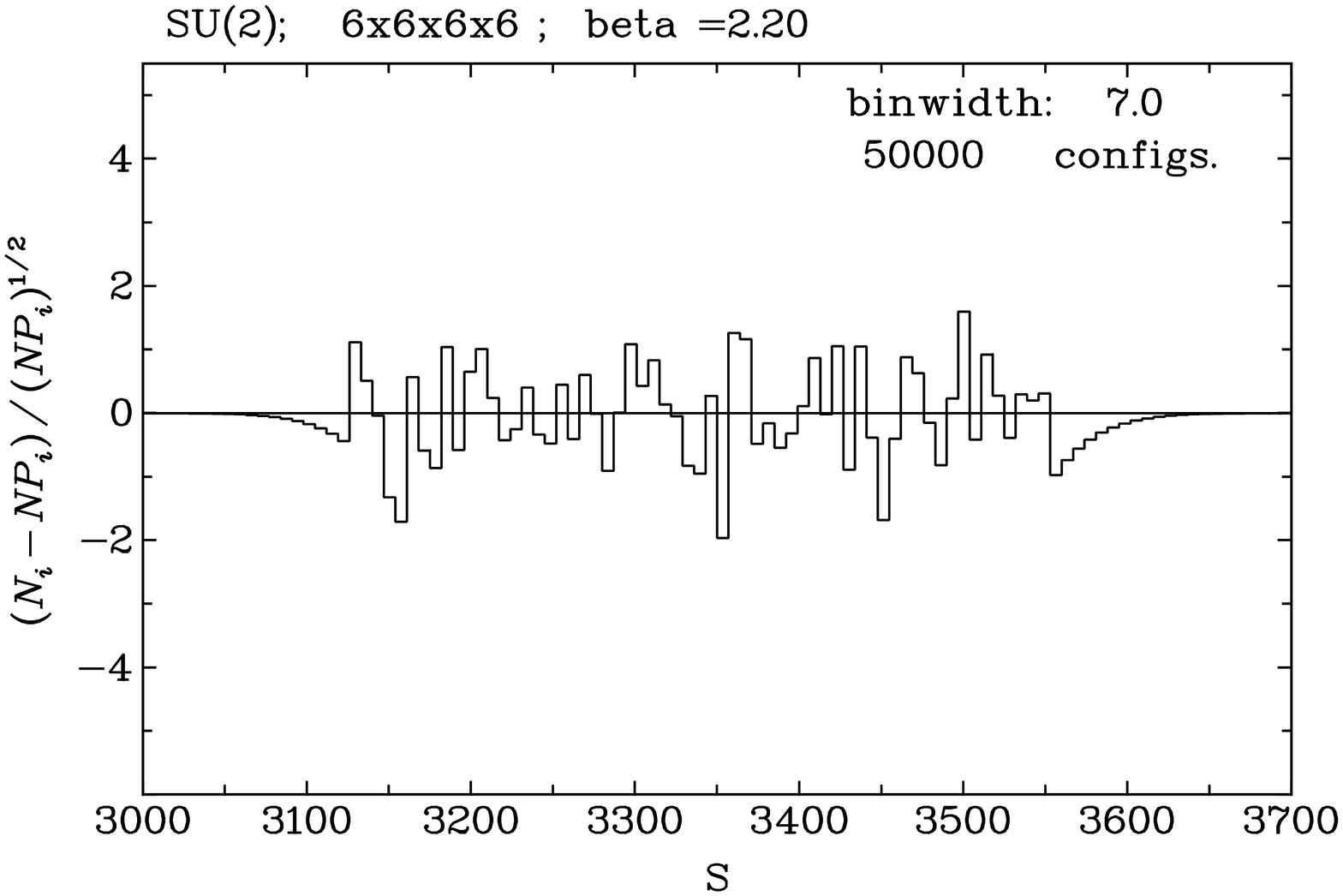}
\caption{$r_i=(N_i-NP_i)/\sqrt{NP_i}$ for $SU(2)$ at $\beta =2.20$ on $4^4$ and $6^4$ lattices.}
\label{fig:fluct}
\end{figure}

\begin{figure}
\includegraphics[width=2in,angle=270]{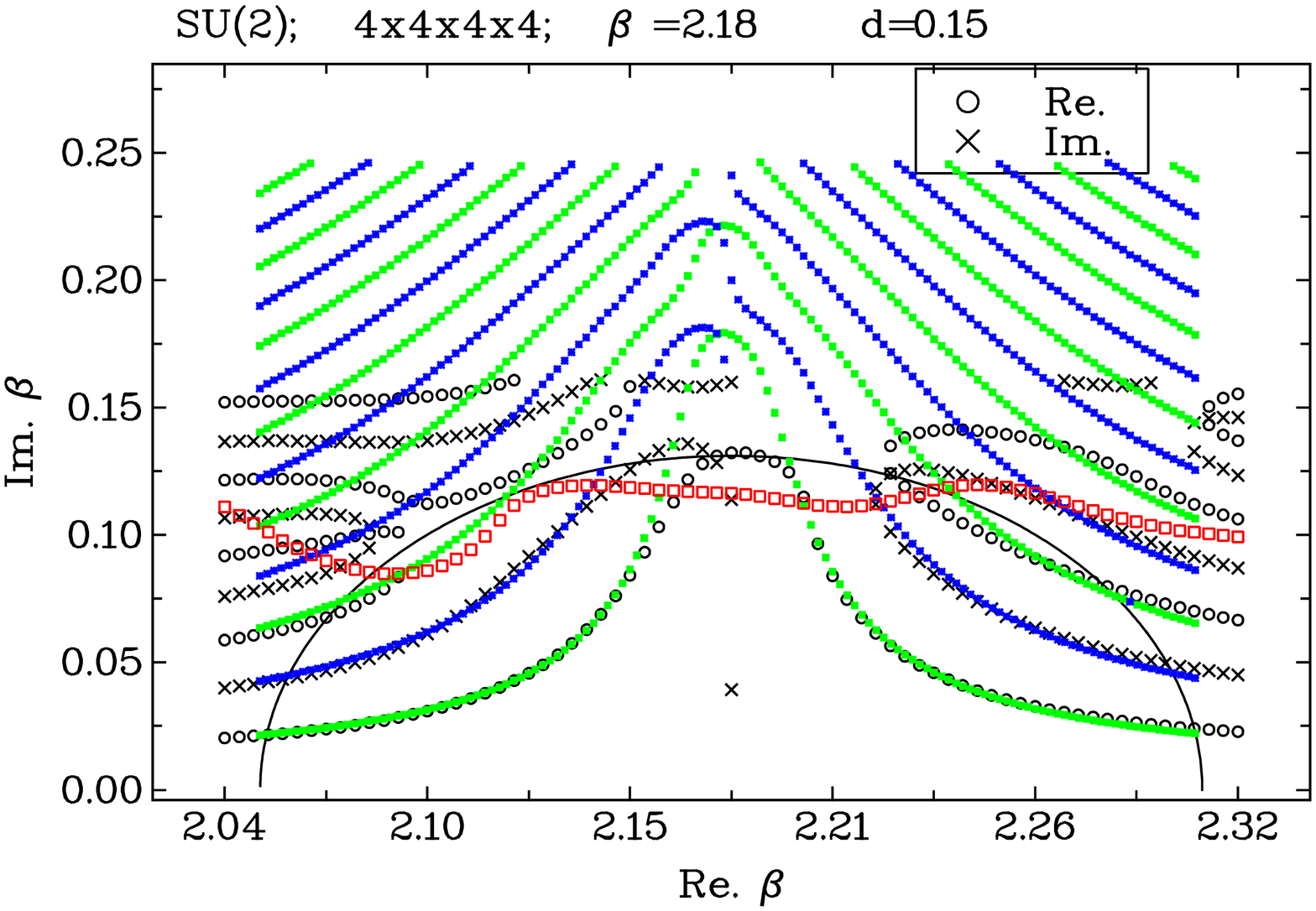}
\includegraphics[width=2in,angle=270]{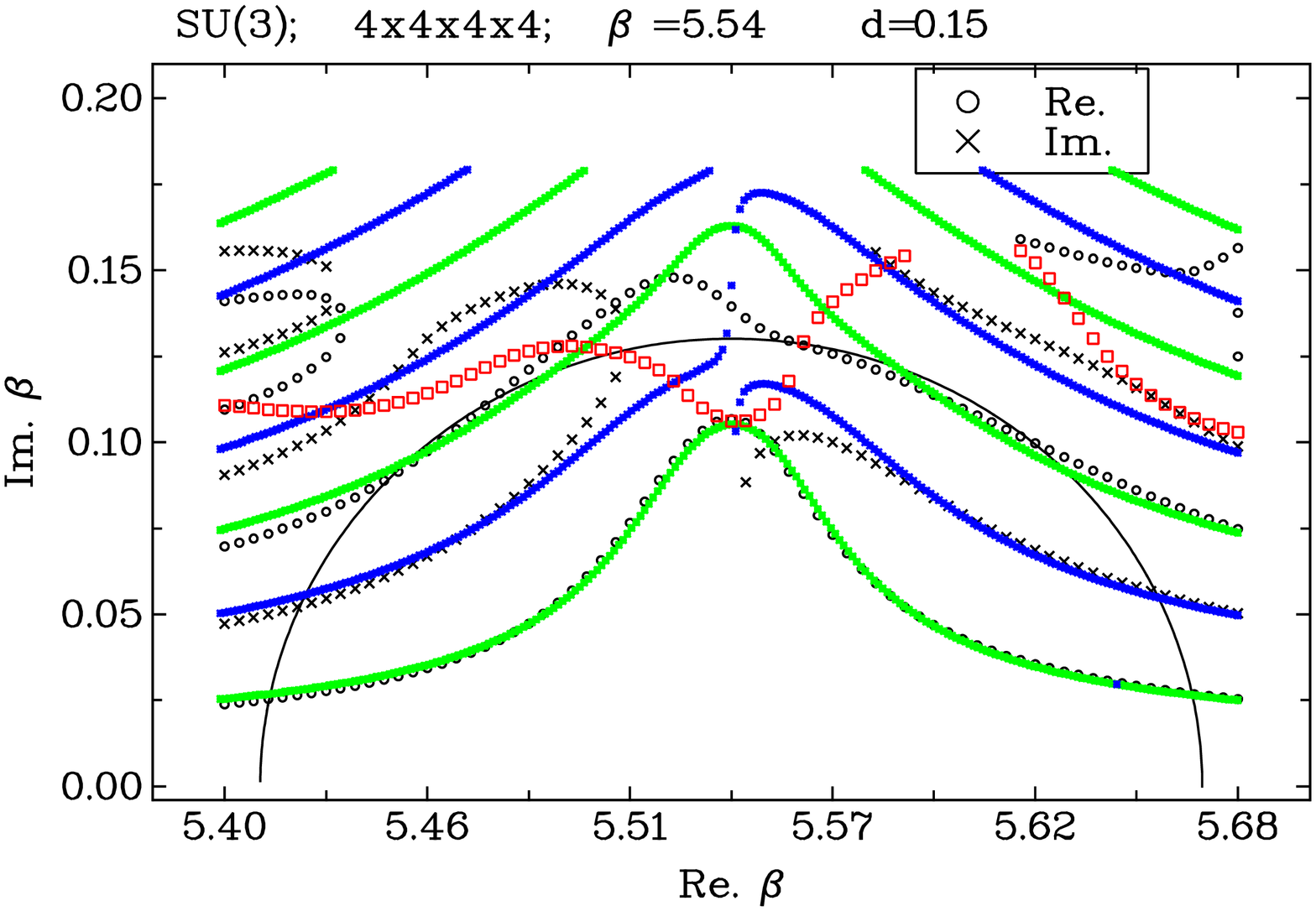}
\caption{Zeros of the real (crosses) and imaginary (circles) using MC on a $4^4$ lattice,
for $SU(2)$  at $\beta$ = 2.18 and $SU(3)$  at $\beta$ =5.54. The smaller dots are the 
values for the real (green) and imaginary (blue) parts 
obtained from the 4 parameter model. The MC exclusion region boundary for $d=0.15$ is represented by boxes (red). 
}
\label{fig:su}
\end{figure}

The zeros come from the deviations from the Gaussian behavior. 
As shown on Fig. \ref{fig:fluct}, discrepancies in unit of the expected fluctuations are coherent for $L=4$ but as the the volume increases, the signal gets lost in the noise (for that particular value of $\beta$). 
The nice regularities of the difference with the Gaussian approximation (for small lattices) suggest 
to fit the distribution with 
\begin{equation}
P(S)\propto \exp(-\lambda_1 S-\lambda_2S^2-\lambda_3 S^3-\lambda_4 S^4)
\label{eq:ps}
\end{equation}
The unknown parameters were determined from the first four moments using Newton's methods 
and also by $\chi^2$ minimization. Very good agreement between the two methods was found on $4^4$ lattices. The zeros can then be calculated for the parametric form (4.2) 
using accurate numerical integration. The results are shown in Fig. \ref{fig:su} on a $4^4$ lattice. 
Comparing results at different $\beta$ on a $4^4$ lattice, we obtained the following locations of the complex zeros:
\begin{itemize}
\item
$\beta =2.18(1) \pm i0.18(2)$ for $SU(2)$  (differs from \cite{falcioni81} 2.23$\pm i0.155$ obtained with MC outside regions of confidence) and another zero at $\beta=2.18(1) \pm i0.22(2)$.
\item
$\beta =5.54(2)\pm i0.10(2)$ for $SU(3)$ (agrees with \cite{alves}) and another zero at $\beta =5.54(2)\pm i0.16(2)$ . 
\end{itemize}
Note that the ratio of the imaginary and real parts of the closest zero is almost 5 times larger in 
$SU(2)$. This indicates that modulations in the perturbative coefficients of $P$ should be easier to see than in $SU(3)$. 
\section{Conclusions}
We have build a  ladder of methods that can be applied for increasing values of the 
imaginary part. 
We found new ways to distinguish fake and true MC zeros that work well with non-Gaussian examples.
Fitting methods based on cubic and quartic perturbations give consistent results at different $\beta$  for larger values of the 
imaginary part on a $4^4$ lattice. Results on larger lattices will be available soon \cite{fisherw}. We are in the process of checking 
the selfconsistency of the parametrization at different $\beta$ and are attempting to extract the 
density of states. 
Effect of an adjoint term, finite-temperature and decimation are also under study.

\acknowledgments
This 
research was supported in part  by the Department of Energy
under Contract No. FG02-91ER40664 and the National Science Foundation 
NSF-PHY-0555693.

\end{document}